\documentclass[copyright,creativecommons]{eptcs} 
\usepackage{graphicx,xcolor}
\usepackage{amsmath}
\usepackage{algorithm}
\usepackage{algpseudocode}
\usepackage{listings}
\lstdefinestyle{cstyle}{
    language=C,
    basicstyle=\ttfamily\small,
    keywordstyle=\color{blue},
    commentstyle=\color{gray},
    stringstyle=\color{green!60!black},
    numbers=left,
    numberstyle=\tiny,
    stepnumber=1,
    numbersep=5pt,
    tabsize=4,
    showstringspaces=false,
    breaklines=true,
    frame=single
}

\title{Verification Challenge: Fractional Cascading for Multi-Nuclide Grid Lookup}
\author{Andrew R.\ Siegel
\institute{Argonne National Laboratory\\Lemont, IL, USA}
\email{siegela@mcs.anl.gov}}
\def\copyrightholders{UChicago Argonne, LLC,\\ Operator of Argonne National Laboratory}
\def\titlerunning{Verification Challenge: Fractional Cascading for Multi-Nuclide Grid Lookup}
\def\authorrunning{Andrew R.\ Siegel}

\begin{document}

\maketitle

\begin{abstract}
We present a verification challenge based on the fractional cascading (FC) technique for accelerating repeated searches across a collection of sorted arrays. The specific context is nuclear cross section lookup in a simulation code, where a material consists of many nuclides, each with its own sorted energy grid. A naive search performs a binary search in each array individually. The FC-based cascade grid structure reduces this cost by performing a single binary search followed by constant-time refinements. The challenge consists of verifying the correctness of the FC algorithm with respect to the naive approach and validating its structural properties.
\end{abstract}

\section{Problem Description}

In particle transport simulations---used to model how particles such as neutrons move through and interact with materials---accurate predictions depend on data known as nuclear cross sections. A nuclear cross section represents the probability of a specific interaction (e.g., scattering or absorption) between a particle and a nucleus, and this probability depends on the particle’s energy.

To compute these interactions,  codes like OpenMC (\url{https://openmc.org}, \cite{ROMANO201590}) store cross section data as numerical values tabulated on a sorted energy grid. Each nuclide (i.e., a specific type of atomic nucleus, such as uranium-235 or hydrogen-1) has its own energy grid, which provides microscopic cross sections—the interaction probabilities for individual nuclei—as a function of incident particle energy (the energy of a particle as it enters the material).

During a simulation, when a particle with a given energy interacts with a material (which may be composed of many nuclides), the code must find where that energy lies in each nuclide’s grid in order to interpolate the corresponding microscopic cross section. This results in repeated searches across multiple similar sorted arrays—one per nuclide.

A straightforward method performs a binary search in each of the $k$ sorted arrays, leading to a total query cost of $O(k\log(n))$, where $n$ is the average size of each grid. This challenge proposes an optimized data structure based on fractional cascading\footnote{\url{https://en.wikipedia.org/wiki/Fractional_cascading}} (FC), which reduces the per-query cost to $O(\log(n) + k)$ by reusing information between searches.

\section{Cascade Grid Algorithm}
The FC technique, summarized in Algorithm \ref{alg:fc}, constructs a cascade grid structure consisting of $k$ augmented energy grids $M_1$, $M_2$, \dots, $M_k$, derived from the original nuclide energy grids $L_1, L_2, \dots, L_k$. Each $M_i$ is a sorted array containing all elements of $L_i$ and every second element of $M_{i+1}$. Each entry $E \in M_i$ is associated with two integer indices: $p_1$ gives the location of $E$ in $L_i$, and $p_2$ gives the approximate index of $E$ in $M_{i+1}$,
accurate up to $\pm{1}$.

The cost of constructing $L_i$ and $M_i$ is fixed and considered amortized over many (typically billions of searches). Once constructed, the search procedure begins with a binary search in $M_1$, then uses index $p_2$ to locate the position in $M_2$, correcting with one comparison if needed, and continues cascading down to $M_k$. Index $p_1$ in each grid provides the lookup location in the corresponding original grid $L_i$.

\begin{algorithm}[H]
  \caption{Cascade Grid Cross Section Lookup}
  \begin{algorithmic}[1]
    \For{$i \gets 1$ to $k$}
      \If{$i = 1$}
        \State $j \gets \text{binary search}(M_i, E)$
      \Else
        \State $j \gets p_2$ (correct if needed by comparing with $M_i[p_2-1]$)
      \EndIf
      \State $p_1, p_2 \gets \text{indices associated with } M_i[j]$
      \State $\sigma_i \gets L_i[p_1]$ \Comment{Microscopic cross section lookup}
    \EndFor
  \end{algorithmic}
  \label{alg:fc}
\end{algorithm}

\section{Implementation and Properties to Verify}

A C implementation of the algorithm is provided in Appendix \ref{app:impl}.  It uses standard data structures and binary search.   A \texttt{main} function performs a small concrete test.  For full verification, one could instead write a driver that initializes a set of sorted arrays, constructs the cascade structure, and compares the results of naive and FC-based lookup routines for a range of energies.

We propose the following properties:
\begin{itemize}
  \item \textbf{Correctness:} For any energy value $E$, the FC-based cascade lookup returns the same indices in $L_1, \dots, L_k$ as $k$ independent binary searches.
  \item \textbf{Structural bound:} The total size of the cascade data structure is bounded above by $2\sum_i |L_i|$.
  \item \textbf{Efficiency (optional):} The lookup procedure performs exactly one binary search and at most one comparison per subsequent grid.
\end{itemize}

\appendix
\section{Implementation}
\label{app:impl}

\begin{lstlisting}[style=cstyle]
#include <stdlib.h>
#include <stdio.h>
#include <math.h>
#include <stdbool.h>

static float *M = NULL;
static int *P1 = NULL, *P2 = NULL;
static int *L_starts = NULL, *M_starts = NULL, *M_sizes = NULL;
static int n_global;

// Binary search on a float array
int binary_search(const float *arr, int size, float key) {
    int lo = 0, hi = size - 1, mid, result = 0;
    while (lo <= hi) {
        mid = (lo + hi) / 2;
        if (arr[mid] <= key) {
            result = mid;
            lo = mid + 1;
        } else {
            hi = mid - 1;
        }
    }
    return result;
}

// Setup the fractional cascading structure
void fc_setup(const float *arrays, const int *sz, int n) {
    n_global = n;
    L_starts = malloc(n * sizeof(int));
    M_starts = malloc(n * sizeof(int));
    M_sizes = malloc(n * sizeof(int));

    int total_L = 0;
    for (int i = 0; i < n; ++i) {
        L_starts[i] = total_L;
        total_L += sz[i];
    }

    int max_total_M = 2 * total_L;
    M = malloc(max_total_M * sizeof(float));
    P1 = malloc(max_total_M * sizeof(int));
    P2 = malloc(max_total_M * sizeof(int));

    int total_M_size = 0;
    for (int i = n - 1; i >= 0; --i) {
        const float *Li = arrays + L_starts[i];
        int Li_size = sz[i];
        int tail_start = (i < n - 1) ? M_starts[i + 1] : 0;
        int tail_size = (i < n - 1) ? (M_sizes[i + 1] + 1) / 2 : 0;

        int Mi_start = total_M_size;
        M_starts[i] = Mi_start;
        int li = 0, ti = 0, mi = 0;

        while (li < Li_size && ti < tail_size) {
            float lv = Li[li];
            float tv = M[tail_start + 2 * ti];
            if (lv <= tv) {
                M[Mi_start + mi] = lv;
                P1[Mi_start + mi] = li;
                P2[Mi_start + mi] = -1;
                li++; mi++;
            } else {
                M[Mi_start + mi] = tv;
                P1[Mi_start + mi] = -1;
                P2[Mi_start + mi] = 2 * ti;
                ti++; mi++;
            }
        }
        while (li < Li_size) {
            M[Mi_start + mi] = Li[li];
            P1[Mi_start + mi] = li;
            P2[Mi_start + mi] = -1;
            li++; mi++;
        }
        while (ti < tail_size) {
            float tv = M[tail_start + 2 * ti];
            M[Mi_start + mi] = tv;
            P1[Mi_start + mi] = -1;
            P2[Mi_start + mi] = 2 * ti;
            ti++; mi++;
        }
        M_sizes[i] = mi;
        total_M_size += mi;
    }
}

int* fc_lookup(const float *arrays, int n, const int *sz, float key) {
    int *indices = malloc(n * sizeof(int));
    int j = binary_search(M + M_starts[0], M_sizes[0], key);

    for (int i = 0; i < n; ++i) {
        int pos = M_starts[i] + j;
        const float *L = arrays + L_starts[i];
        int L_size = sz[i];
        if (P1[pos] != -1) {
            // If key is beyond the end of the array, return last index
            if (key >= L[L_size - 1]) {
                indices[i] = L_size - 1;
            } else {
                indices[i] = P1[pos];
            }
        } else {
            int idx = 0;
            while (idx + 1 < L_size && L[idx + 1] <= key) idx++;
            indices[i] = idx;
        }

        if (i < n - 1) {
            int guess = P2[pos];
            int next_start = M_starts[i + 1];
            int next_size = M_sizes[i + 1];

            if (guess >= 1 && guess < next_size &&
                M[next_start + guess] > key &&
                M[next_start + guess - 1] <= key) {
                j = guess - 1;
            } else if (guess >= 0 && guess < next_size) {
                j = guess;
            } else {
                // Fallback: re-search if guess is invalid
                j = binary_search(M + next_start, next_size, key);
            }
        }
    }
    return indices;
}

// Naive lookup to validate fc_lookup
int* naive_lookup(const float *arrays, const int *sz, int n, float key) {
    int *indices = malloc(n * sizeof(int));
    int offset = 0;
    for (int i = 0; i < n; ++i) {
        int s = sz[i];
        int idx = 0;
        while (idx + 1 < s && arrays[offset + idx + 1] <= key)
            idx++;
        indices[i] = idx;
        offset += s;
    }
    return indices;
}

int main(void) {
    int n = 3;
    int sz[] = {5, 6, 4};
    float arrays[] = {
        1.0, 2.0, 3.0, 4.0, 5.0,
        1.5, 2.5, 3.5, 4.5, 5.5, 6.5,
        0.5, 1.5, 2.5, 3.5
    };

    // Setup the fractional cascade structure
    fc_setup(arrays, sz, n);

    // Run multiple tests
    float test_keys[7] = {0.0, 1.4, 2.0, 3.2, 4.7, 6.0, 7.0};
    int num_tests = sizeof(test_keys) / sizeof(test_keys[0]);

    for (int t = 0; t < num_tests; ++t) {
        float key = test_keys[t];
        int *fc_result = fc_lookup(arrays, n, sz, key);
        int *naive_result = naive_lookup(arrays, sz, n, key);

        printf("Key = %.2f\n", key);
        bool ok = true;
        for (int i = 0; i < n; ++i) {
            printf("  Array %d: fc = %d, expected = %d\n",
                   i, fc_result[i], naive_result[i]);
            if (fc_result[i] != naive_result[i])
                ok = false;
        }
        if (!ok)
            printf("MISMATCH DETECTED\n");
        else
            printf("Match\n");
        printf("\n");

        free(fc_result);
        free(naive_result);
    }

    // Clean up
    free(M);
    free(P1);
    free(P2);
    free(L_starts);
    free(M_starts);
    free(M_sizes);
}
\end{lstlisting}

\paragraph{Acknowledgment.}
This publication is based on work supported by the U.S. Department of Energy under contract DE-AC02-06CH11357.

\bibliographystyle{eptcs}
\bibliography{fc}

\begin{thebibliography}{1}
\providecommand{\bibitemdeclare}[2]{}
\providecommand{\surnamestart}{}
\providecommand{\surnameend}{}
\providecommand{\urlprefix}{Available at }
\providecommand{\url}[1]{\texttt{#1}}
\providecommand{\href}[2]{\texttt{#2}}
\providecommand{\urlalt}[2]{\href{#1}{#2}}
\providecommand{\doi}[1]{doi:\urlalt{https://doi.org/#1}{#1}}
\providecommand{\eprint}[1]{arXiv:\urlalt{https://arxiv.org/abs/#1}{#1}}
\providecommand{\bibinfo}[2]{#2}

\bibitemdeclare{article}{ROMANO201590}
\bibitem{ROMANO201590}
\bibinfo{author}{Paul~K. \surnamestart Romano\surnameend},
  \bibinfo{author}{Nicholas~E. \surnamestart Horelik\surnameend},
  \bibinfo{author}{Bryan~R. \surnamestart Herman\surnameend},
  \bibinfo{author}{Adam~G. \surnamestart Nelson\surnameend},
  \bibinfo{author}{Benoit \surnamestart Forget\surnameend} \&
  \bibinfo{author}{Kord \surnamestart Smith\surnameend} (\bibinfo{year}{2015}):
  \emph{\bibinfo{title}{{OpenMC}: A state-of-the-art {M}onte {C}arlo code for
  research and development}}.
\newblock {\slshape \bibinfo{journal}{Annals of Nuclear Energy}}
  \bibinfo{volume}{82}, pp. \bibinfo{pages}{90--97},
  \doi{10.1016/j.anucene.2014.07.048}.
\newblock \bibinfo{note}{Joint International Conference on Supercomputing in
  Nuclear Applications and Monte Carlo 2013, SNA + MC 2013. Pluri- and
  Trans-disciplinarity, Towards New Modeling and Numerical Simulation
  Paradigms}.

\end{thebibliography}

\end{document}